\documentclass[twocolumn, showpacs]{revtex4}
\usepackage{graphicx}

\begin{document}

\title{States of fermionic atoms in an optical superlattice across a Feshbach
resonance}
\author{T. Goodman and L.-M. Duan}
\affiliation{FOCUS center and MCTP, Department of Physics,
University of Michigan, Ann Arbor, MI 48109} 

\begin{abstract}
We investigate states of fermionic atoms across a broad Feshbach
resonance in an optical superlattice which allows interaction only
among a small number of lattice sites. The states are in general
described by superpositions of atomic resonating valence bonds and
dressed molecules. As one scans the magnetic field, level crossing
is found between states with different symmetry properties, which
may correspond to a quantum phase transition in the many-body
case. 
\end{abstract}

\maketitle

\section{Introduction}

The past several years have seen many new developments in ultracold atomic
physics \cite{advances1, advances2}. Two experimental control techniques
have played a critical role in these advances: the use of optical lattices
to produce diverse interaction configurations \cite{optlat}, and Feshbach
resonance to control the strength of the interactions between atoms \cite
{feshbach}. This has motivated significant interest in combining these two
techniques \cite{latfesh1,duan, latfesh2, ho, latfesh4, latfesh5}.

Close to a broad Feshbach resonance, one then has strongly interacting
fermions in a lattice, with the effective interaction energy easily larger
than the band gap. For such a strongly correlated system, in general it is
very hard to understand its physics. As the first step, one may try to
understand the single-site physics, which contains the dominant interactions
for the atoms in a deep optical lattice. The two-body physics on a single
site has been solved exactly in \cite{ho} under typical experimental
configurations, and it is found there that many bands of the lattice are
significantly populated, in agreement with the experimental observation \cite
{latfesh4}. In the next step, it is natural to consider interactions over a
few lattice sites. We have shown in \cite{duan} that for the case of a
multiple-site lattice, one needs to take into account the direct collision
interactions between the neighboring sites, as the magnitude of this
interaction is larger than the atom tunneling rate for a typical wide
Feshbach resonance such as with $^{40}K$ or $^{6}Li$ atoms. An effective
lattice interaction Hamiltonian has been derived there, which takes into
account both the multi-band populations and the direct off-site and
on-site collisions. In this effective Hamiltonian, whenever two atoms come
to the same lattice site, they form a dressed molecule state (a single-site
Cooper pair), which corresponds to an exact eigenstate constructed from the
single-site physics. The effective Hamiltonian then describes the
interaction between these dressed molecules and the atoms over different
lattice sites. The explicit form of the Hamiltonian is as follows \cite{duan}

\begin{eqnarray}
H_{eff} &=&\sum_{i}\Delta (B)d_{i}^{\dagger }d_{i}+\sum_{i;j\in
N(i)}t_{d}Pd_{i}^{\dagger }d_{j}P  \nonumber \\
{} &{}&+\sum_{i;j\in N(i)}\sum_{\sigma }\left( t_{a}Pa_{i\sigma }^{\dagger
}a_{j\sigma }P+t_{da}d_{i}^{\dagger }d_{j}a_{j\sigma }^{\dagger }a_{i\sigma
}\right)   \nonumber \\
{} &{}&+\sum_{i;j\in N(i)}\left( gd_{i}^{\dagger }\left( a_{i\uparrow
}a_{j\downarrow }-a_{i\downarrow }a_{j\uparrow }\right) +h.c.\right)
\end{eqnarray}
where $a^{\dagger }$ and $d^{\dagger }$ represent creation operators for
fermionic atoms and bosonic dressed molecules, respectively; $\sigma
=\uparrow ,\downarrow $ labels two internal spin states; and $i$ labels the
lattice sites (with $j\in N(i)$ labeling the sites adjacent to $i$). $P$
represents a projection of the state at every lattice site $i$ onto the
four-dimensional subspace with basis $\left\{ \left| 0\right\rangle
_{i},a_{i\uparrow }^{\dagger }\left| 0\right\rangle _{i},a_{i\downarrow
}^{\dagger }\left| 0\right\rangle _{i},d_{i}^{\dagger }\left| 0\right\rangle
_{i}\right\} $. The parameters $g$, $t_{d}$, $t_{a}$, and $t_{da}$ depend on
the physical properties of the atomic system as well as the multi-band
properties of the optical lattice (see the explicit expressions in \cite
{duan}). The parameter $\Delta $ corresponds to an eigenenergy of the
two-body physics from a single site, and can be tuned over a wide range of
values by varying the applied magnetic field $B$ \cite{ho}.

In this paper, we make use of the above effective Hamiltonian
$H_{eff}$ to study the physics of this strongly correlated system
with interactions among a few lattice sites. In particular, we
focus on investigation of the states of a single plaquette, which
is a basic unit of the two-dimensional square lattice. This study
has two purposes. First, understanding the states of atoms at a
single plaquette is a necessary step towards the challenging goal
of understanding the physics of this strongly interacting gas in a
quasi-two-dimensional optical lattice. It is shown in \cite{duan}
that the effective Hamiltonian $H_{eff}$ reduces\ to the
well-known t-J model \cite{tJmod} and the XXZ model \cite{xxz} in
certain limits of the parameter values, so the physics associated
with $H_{eff}$ should be rich and parameter-dependent. In this
work, we would like to understand the influence of the parameters
on a few body physics, and that understanding will provide an
intuition for taking appropriate approximations towards the
many-body physics. In particular, from the few-body physics, we
construct states which will provide the basic entries for an
effective many-body theory through the contractor renormalization
procedure (a real-space renormalizaton group method for high
dimensions) \cite{core1,core2}. We will see that even for a single
plaquette, the behavior of the states has been pretty rich. The
eigenstates involve resonating valence bonds and superposition of
dressed molecules, and are highly entangled over different lattice
sites. With variation of the parameters in the Hamiltonian
$H_{eff}$, there are several level crossings for the lowest
eigenstate with change of the state symmetry properties, which may
correspond to a quantum phase transition for larger systems.

\begin{figure}[tbp]
\includegraphics[width=229.5pt]{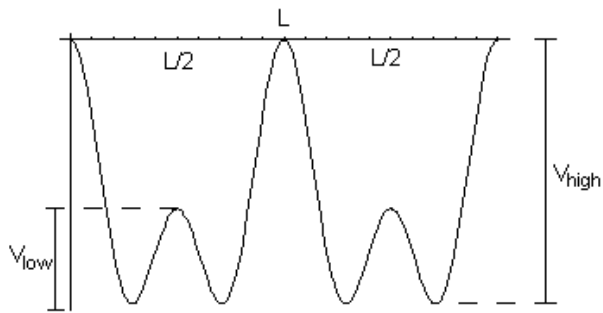}
\caption{Superlattice potential vs. $x$ for $V(x) = - \left( V_1 sin^2
\left( \frac{\protect\pi x}{L} \right) + V_2 sin^2 \left( \frac{2 \protect\pi
x}{L} \right) \right)$}
\label{superlat}
\end{figure}

Second, the study of the physics of a single plaquette is also of
practical relevance. For atoms in an optical superlattice, the
physics can be dominated by the interactions within single
plaquettes. A simple optical superlattice can be formed by adding
two standing wave laser beams with commensurate wave vectors
\cite{sl1,sl2}. The potential, say, along the $x$ direction, has
the form $V(x)=-\left[ V_{1}sin^{2}\left( k_{1}x\right)
+V_{2}sin^{2}\left( k_{2}x\right) \right] $. If we choose the wave
vector $k_{2}=2k_{1}=2 \pi /L$ and require $0< V_1 <4V_2$, then we
have potential barriers of two different heights (see
FIG. \ref{superlat}). The minima occur at $x=nL \pm x_0$ (for
integer $n$), where
\begin{equation}
x_0 = \frac{L}{2\pi}cos^{-1}\left(\frac{-V_1}{4 V_2}\right)
\end{equation}
The lower and higher potential barriers $V_{low}$ and $V_{high}$ are given
respectively by
\begin{eqnarray}
V_{low} &=&\left( 1-\frac{V_{1}}{4V_{2}}\right) ^{2}V_{2}\nonumber \\
V_{high} &=&\left( 1+\frac{V_{1}}{4V_{2}}\right) ^{2}V_{2}.
\end{eqnarray}
The barrier $V_{high}$ can be significantly larger than $V_{low}$ if we
choose $V_{1}$ close to $4V_{2}$, and such a high barrier turns off the
interactions except for the ones between the sites separated by $V_{low}$.
If we apply this optical superlattice potential along both the x and y
directions and a deep lattice potential along the z direction, we then have
interactions dominantly within the single plaquettes in the x-y plane. With
strongly interacting atoms in this optical superlattice potential, one can
test the predictions from the effective Hamiltonian $H_{eff}$, and detect
the exotic entangled states emerging from the ground state configurations of
$H_{eff}$.

We should also point out that the effective Hamiltonian $H_{eff}$ includes
the Hubbard model as a particular case. The Hubbard model is given by the
Hamiltonian \cite{core2,hm}
\begin{equation}
H_{Hub}=-t\sum_{<i,j>,\sigma }{\left( a_{i\sigma }^{\dagger }a_{j\sigma
}+H.c.\right) }+U\sum_{i}{n_{i\uparrow }n_{i\downarrow },}
\end{equation}
where $n_{i\sigma }=a_{i\sigma }^{\dagger }a_{i\sigma }$. Specifically,
$H_{eff}$ can be written in the form of $H_{Hub}$ if we substitute $%
d_{i}^{\dagger }$ with $a_{i\uparrow }^{\dagger }a_{i\downarrow }^{\dagger }$
and make a particular choice of the parameters in $H_{eff}$ with $t_{a}=-t$,
$t_{da}=t$, $g=t$, $t_{d}=0$, and $\Delta =U$. So, one can see that $H_{eff}$
extends the well-known Hubbard model $H_{Hub}$ in a nontrivial way. Note
that for strongly interacting atoms near a broad Feshbach resonance, the
parameters $g$ and $t_{da}$ are significantly different from the atomic
tunneling rate $t$ due to the multi-band populations and the direct
neighboring collisions. From the expressions of these parameters in Ref.
\cite{duan}, we estimate that typically $\left| t_{d}\right| <<\left|
t_{a}\right| <<\left| t_{da}\right| \sim $ $|g|$. This is because $t_{a}$
corresponds to atomic tunneling in the single lowest band, whereas $t_{da}$
and $g$ correspond to interactions involving the dressed molecule states
(which are superpositions of states in multiple upper bands). For the
numerical calculations in this work, we typically take $t_{d}\sim 0,$
$-t_{a}\sim $ $0.1|g|$--$0.3|g|$, and $t_{da}\sim $ $|g|$--$2|g|$. (Note
from the form of $H_{eff}$ that the sign of $g$ is essentially irrelevant,
as it can be incorporated into the definition of $d^\dagger$.)  The
parameter $\Delta $ is sensitive to the external magnetic field, and can
be scanned from the value much smaller than $-\left| g\right| $ to the value
much larger than $\left| g\right| $.

The atomic states within each plaquette critically depend on the atom number
and the spin configuration in that plaquette. In the following, we will
consider all the different nontrivial cases with different numbers of spin $%
\uparrow $ and $\downarrow $ atoms occupying the four-site plaquette.

\section{Four atoms per plaquette: two $\uparrow $, two $\downarrow $}

Over most of the typical range of the parameter values, the plaquette
occupied by two $\uparrow $ and two $\downarrow $ atoms has two distinct
types of ground states, with a level crossing occurring at some critical
value of $\Delta $. These two types of states can be distinguished by how
they transform under a $90^{\circ }$ rotation in the plane of
the plaquette. Under such a rotation, the ground state wavefunction for $%
\Delta $ less than (greater than) the critical value is multiplied by a
factor of $+1$ ($-1$). Thus, we say that the phase on the negative side of
the transition has s-wave symmetry, and that on the positive side has d-wave
symmetry.

\begin{figure}[tbp]
\includegraphics[width=229.5pt]{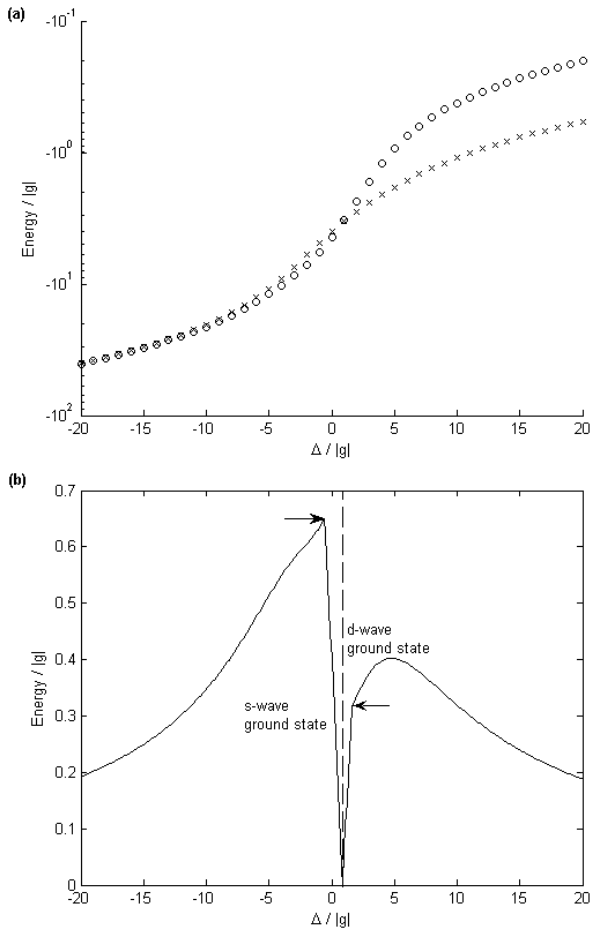}
\caption{Energy vs. $\Delta$ for a plaquette occupied by two $\uparrow$ and
two $\downarrow$ atoms. Other parameters are $t_{da} = 1.5 |g|$, $t_a = -0.2
|g|$, $t_d = 0$ (a): Eigenenergies of the s-wave ($\circ$) and d-wave (x)
states. (b): Energy difference (gap) between ground state and first excited
state. The gap vanishes at the level crossing point. Because the
eigenenergies vary smoothly with $\Delta$, the curve is smooth except at the
level crossing points for the ground state (where the gap is zero) and for
the first excited state (indicated by arrows)}
\label{spect22}
\end{figure}

The ground states of each of these two types change smoothly with changes in
the parameter $\Delta $. Thus, we can identify particular energy eigenstates
as the ``s-wave state'' and the ``d-wave state'' over the full range of $%
\Delta $, even as the exact form of the eigenstate changes. (Note that these
are not the \textit{only} eigenstates with s-wave and d-wave symmetry --
here we use these terms to refer solely to those states which are the
ground states when the system is in the corresponding parameter regions.) The
energies of the s-wave and d-wave states can be easily calculated through
exact diagonalization, and they are plotted in FIG. \ref{spect22}(a), which
illustrates the crossover between them. (For this figure, we scan $\Delta $
and set the other parameters of $H_{eff}$ to their typical values with $%
t_{da}=1.5|g|$ and $t_{a}=-0.2|g|$.) The energy gap between the ground state
and 1st excited state is shown in FIG. \ref{spect22}(b)).

To understand the properties of the ground state, it is important to have
its explicit expression. Although one can calculate this explicit expression
through numerical exact diagonalization, the state is in general a
superposition of many basis-vectors ($36$ vectors in this case), with all
the superposition coefficients varying with $\Delta $. It is troublesome to
understand the state's properties from this lengthy expression. To overcome
this problem, we describe the s-wave and d-wave states more compactly, in a
way that illustrates their rotational symmetry, by means of a pictorial
representation which we define here. The four sites of a plaquette we label
as: \includegraphics{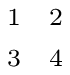}. We place various pictures on these sites
corresponding to creation operators applied at those sites. The whole
picture represents the product of these operators, applied to the vacuum
state $\left| 0\right\rangle $. For instance, \includegraphics{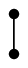}
placed on two sites (either horizontally, vertically, or diagonally)
represents a normalized singlet between those two sites. So, if the sites
are labeled $i$ and $j$, this represents $\frac{1}{\sqrt{2}}\left(
a_{i\uparrow }^{\dagger }a_{j\downarrow }^{\dagger }-a_{i\downarrow
}^{\dagger }a_{j\uparrow }^{\dagger }\right) $. (Note that the order of $i$
and $j$ does not matter, as the anti-commutation of $a_{i}^{\dagger }$ and $%
a_{j}^{\dagger }$ makes the singlet symmetric under exchange of $i$ and $j$%
.) \includegraphics{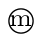} represents a dressed molecule. (I.e., if
located at site $i$ , this picture corresponds to $d_{i}^{\dagger }$.) %
\includegraphics{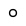} represents an unoccupied site. The creation
operators that make up a singlet are always grouped together; other than
that, the order of the operators is irrelevant, as the singlets and dressed
molecules commute. As an example, the picture %
\includegraphics{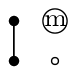} represents $\frac{1}{\sqrt{2}}\left(
a_{1\uparrow }^{\dagger }a_{3\downarrow }^{\dagger }-a_{1\downarrow
}^{\dagger }a_{3\uparrow }^{\dagger }\right) d_{2}^{\dagger }\left|
0\right\rangle $.

The full Hilbert space for two $\uparrow $ and two $\downarrow $ atoms on a
plaquette is $36$-dimensional. However, the s-wave state (over the full
range of $\Delta $) can be conveniently expressed as a vector in a
4-dimensional subspace of the full space, with basis vectors:

\includegraphics{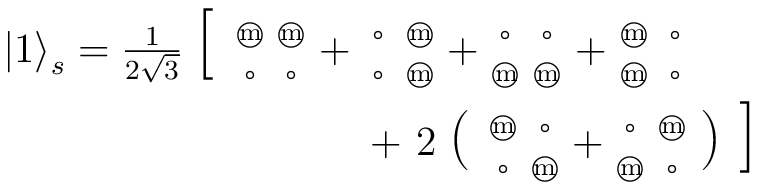}

\includegraphics{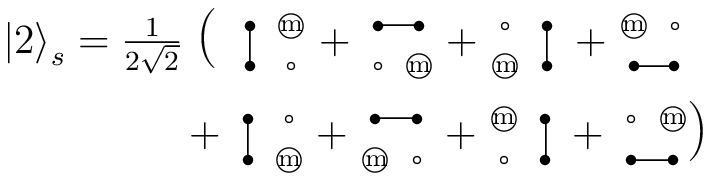}

\includegraphics{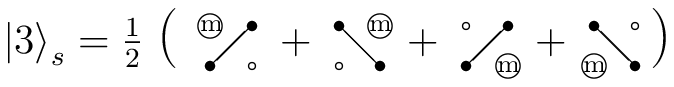}

\includegraphics{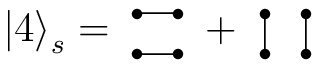}

The d-wave state (as $\Delta $ varies) can be written as a vector in a
3-dimensional subspace with basis vectors:

\includegraphics{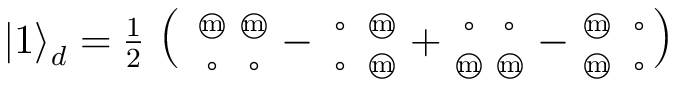}

\includegraphics{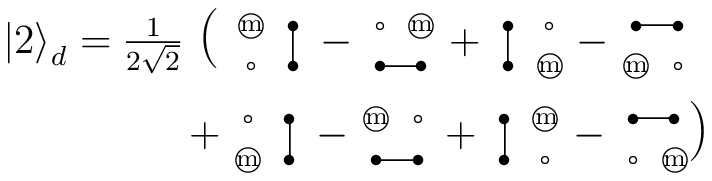}

\includegraphics{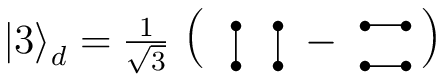}

Note that $\left| 4\right\rangle _{s}$ and $\left| 3\right\rangle _{d}$ are
written here as sums of \textit{non-orthogonal} terms; however, this form
makes their rotational symmetry readily apparent. The states $\left|
4\right\rangle _{s}$ and $\left| 3\right\rangle _{d}$ are the resonating
valence bond (RVB) states for atoms on a single plaquette \cite{tJmod,core2}%
, with s and d wave symmetries, respectively. For a larger lattice, the RVB\
states are in general superpositions of many different spin-singlet
distribution patterns \cite{rvb}.

\begin{figure}[tbp]
\includegraphics[width=229.5pt]{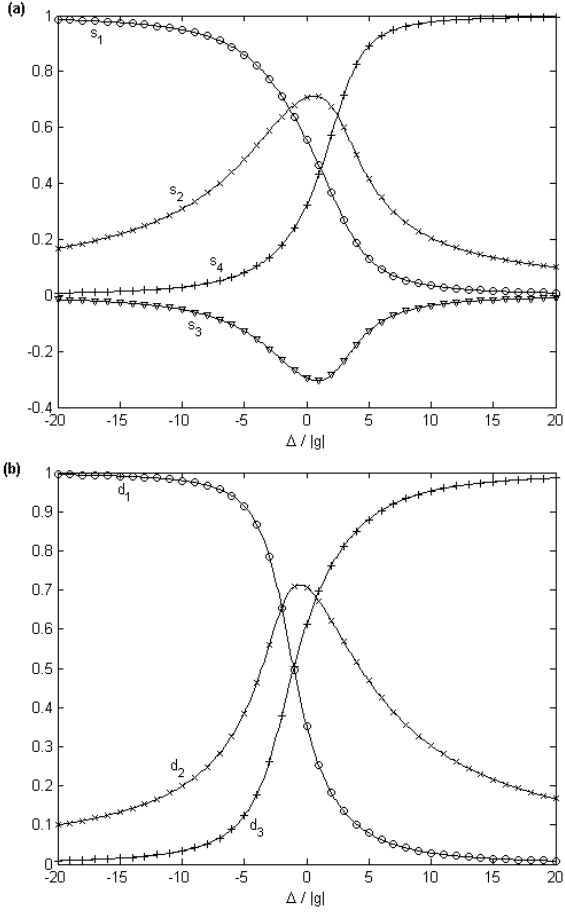}
\caption{The ground-state configuration vs. the detuning $\Delta $ for a
plaquette occupied by two $\uparrow $ and two $\downarrow $ atoms. ($%
t_{da}=1.5|g|$, $t_{a}=-0.2|g|$, $t_{d}=0$) (a) Components of the s-wave state
($s_{1}$: $\circ $, $s_{2}$: x, $s_{3}$: {\protect\footnotesize $%
\bigtriangledown $}{\protect\small , $s_{4}:$ +). (b) Components of the
d-wave state ($d_{1}$: $\circ $, $d_{2}$: x, $d_{3}$: +). The marked data
points were computed from the full Hamiltonian $H_{eff}$, whereas the solid
lines were computed from the projected Hamiltonians $H_{s}$ and $H_{d}$,
respectively.}}
\label{coeff22}
\end{figure}

Thus the s-wave and d-wave ground states, respectively, are:
\begin{eqnarray}
\left| \psi \right\rangle _{s} &=&s_{1}\left| 1\right\rangle
_{s}+s_{2}\left| 2\right\rangle _{s}+s_{3}\left| 3\right\rangle
_{s}+s_{4}\left| 4\right\rangle _{s}, \\
\left| \psi \right\rangle _{d} &=&d_{1}\left| 1\right\rangle
_{d}+d_{2}\left| 2\right\rangle _{d}+d_{3}\left| 3\right\rangle _{d},
\end{eqnarray}
They are superpositions of many different distribution patterns of the
dressed molecules and the atomic valence bonds (spin singlets). The values
of the superposition coefficients are shown in FIG. \ref{coeff22} as a
function of the ratio $\Delta /|g|$. Note that in the limiting case $\Delta
/|g|>>1$, the effective Hamiltonian $H_{eff}$ reduces to the t-J model \cite
{duan}. Indeed, one can see from FIG. \ref{coeff22} that the state $\left|
\psi \right\rangle _{d}$ in that limit tends to the ground state $\left|
3\right\rangle _{d}$ of the t-J model on a plaquette \cite{tJmod}.

Projected onto these subspaces, $H_{eff}$ (with $t_{d}\simeq 0$) expressed
in terms of the bases shown above becomes:

{\small
\begin{equation}
H_{s}=\left(
\begin{array}{cccc}
2\Delta & -2\sqrt{3}g & 0 & 0 \\
-2\sqrt{3}g & \Delta & \sqrt{2}\left( t_{a}+t_{da}\right) & -2g \\
0 & \sqrt{2}\left( t_{a}+t_{da}\right) & \Delta & 0 \\
0 & -2g & 0 & 0
\end{array}
\right)
\end{equation}
}for the s-wave state, and:{\normalsize \ }
\begin{equation}
H_{d}=\left(
\begin{array}{ccc}
2\Delta & -2g & 0 \\
-2g & \Delta & -2\sqrt{3}g \\
0 & -2\sqrt{3}g & 0
\end{array}
\right)
\end{equation}
for the d-wave state. The lowest energy eigenstates of these two
Hamiltonians are the s-wave and d-wave states (respectively) of the full
Hamiltonian $H_{eff}$. (See solid lines on FIG. \ref{coeff22}.)

\begin{figure}[tbp]
{\small \includegraphics[width=229.5pt]{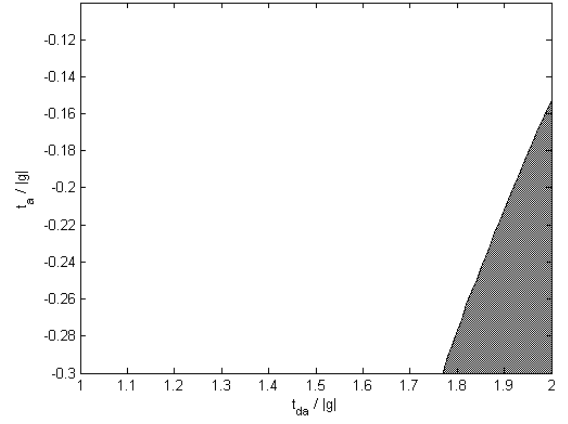} }
\caption{In the typical range of $t_{da}/|g|$ and $t_{a}/|g|$, the s-wave
triplet ground state of the plaquette with 2 $\uparrow $ and 2 $\downarrow $
atoms occurs for parameter values within the shaded region.}
\label{triprange}
\end{figure}

\begin{figure}[tbp]
{\small \includegraphics[width=229.5pt]{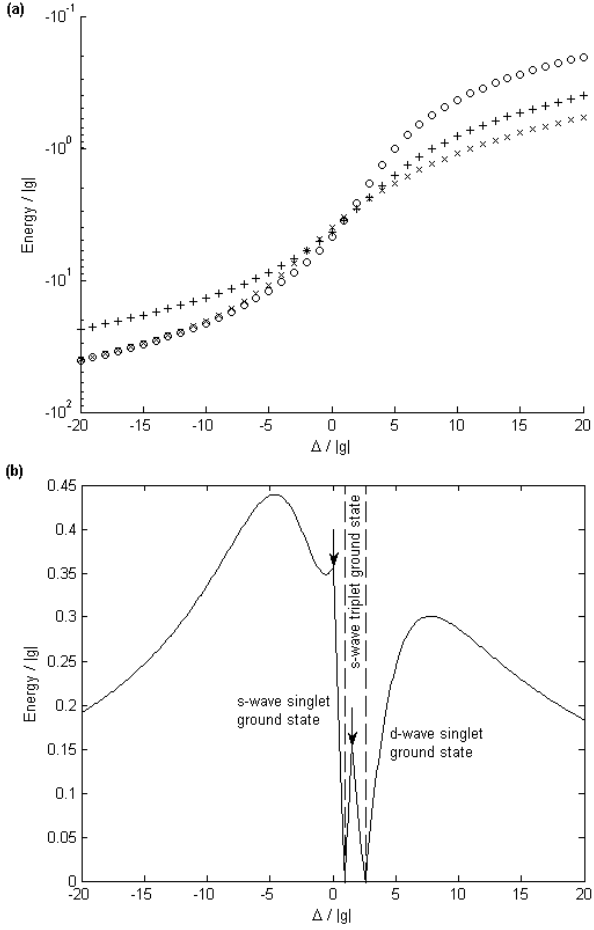} }
\caption{Energy vs. $\Delta $ for a plaquette occupied by two $\uparrow $
and two $\downarrow $ atoms. Other parameters are $t_{da}=2|g|$, $t_{a}=
-0.3|g|$, $t_{d}=0$ (a): Eigenenergies of the s-wave singlet ($\circ $),
d-wave singlet (x), and s-wave triplet (+) states. (b): Energy difference
between ground state and first excited state. Crossovers in the first excited
state are indicated by arrows.}
\label{spect22trip}
\end{figure}

For a small portion of the range of the parameter values $t_{da}/|g|$ and
$t_{a}/|g|$ there is an additional type of ground state, which occurs for $%
\Delta $ between the s-wave and d-wave states above. For this type, the
ground state also has s-wave rotational symmetry. However, the states $%
\uparrow $ and $\downarrow $ are in a triplet configuration, rather than the
singlet occurring in the other two types of states $\left| \psi
\right\rangle _{s}$ and $\left| \psi \right\rangle _{d}$. The region of the
parameter space for which this triplet phase occurs is shown in FIG. \ref
{triprange}. The eigenenergies of the s-wave singlet, s-wave triplet, and
d-wave singlet states are shown in FIG. \ref{spect22trip}(a) for $t_{da}=
2|g|$, $t_{a}=-0.3|g|$ (which is within the range where the triplet
ground state occurs.) The gap between the ground state and first excited state
for $t_{da}=2|g|$, $t_{a}=-0.3|g|$ is shown in FIG. \ref{spect22trip}(b).

The s-wave triplet state can be written as a linear combination of three
states:

\includegraphics{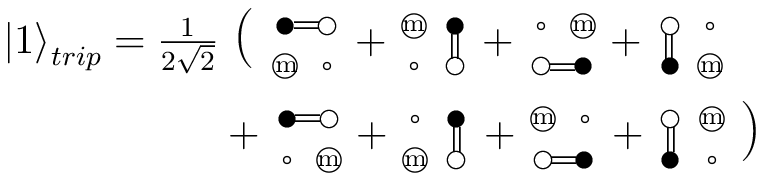}

\includegraphics{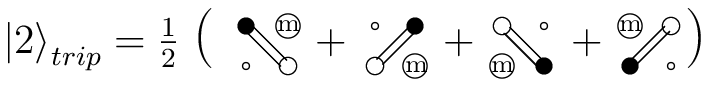}

\includegraphics{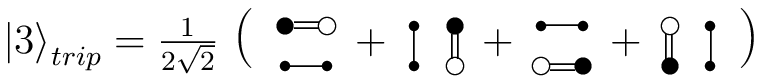}

Here \includegraphics{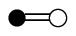} represents the triplet $\frac{1}{\sqrt{2%
}}\left( a_{i\uparrow }^{\dagger }a_{j\downarrow }^{\dagger }+a_{i\downarrow
}^{\dagger }a_{j\uparrow }^{\dagger }\right) $, where $i$ is the site of the
black-filled circle, and $j$ is the site of the white-filled circle. Note
that unlike the singlet, the triplet is not symmetric under exchange of $i$
and $j$: $\frac{1}{\sqrt{2}}\left( a_{i\uparrow }^{\dagger }a_{j\downarrow
}^{\dagger }+a_{i\downarrow }^{\dagger }a_{j\uparrow }^{\dagger }\right) =-%
\frac{1}{\sqrt{2}}\left( a_{j\uparrow }^{\dagger }a_{i\downarrow }^{\dagger
}+a_{j\downarrow }^{\dagger }a_{i\uparrow }^{\dagger }\right) $

Projected onto the basis $\left\{ \left| 1\right\rangle _{trip},\left|
2\right\rangle _{trip},\left| 3\right\rangle _{trip}\right\} $, $H_{eff}$
(with $t_{d}\simeq 0$) becomes:

{\small
\begin{equation}
H_{trip}=\left(
\begin{array}{ccc}
\Delta & \sqrt{2}\left( t_{da}-t_{a}\right) & -2\sqrt{2}g \\
\sqrt{2}\left( t_{da}-t_{a}\right) & \Delta & 0 \\
-2\sqrt{2}g & 0 & 0
\end{array}
\right)
\end{equation}
}and the ground state of this Hamiltonian is the s-wave triplet state. It
should also be noted that the s-wave triplet state is the first excited
state of $H_{eff}$ in the limit of large positive $\Delta $.

\section{Four atoms per plaquette: three $\uparrow $, one $\downarrow $}

\begin{figure}[tbp]
{\small \includegraphics[width=229.5pt]{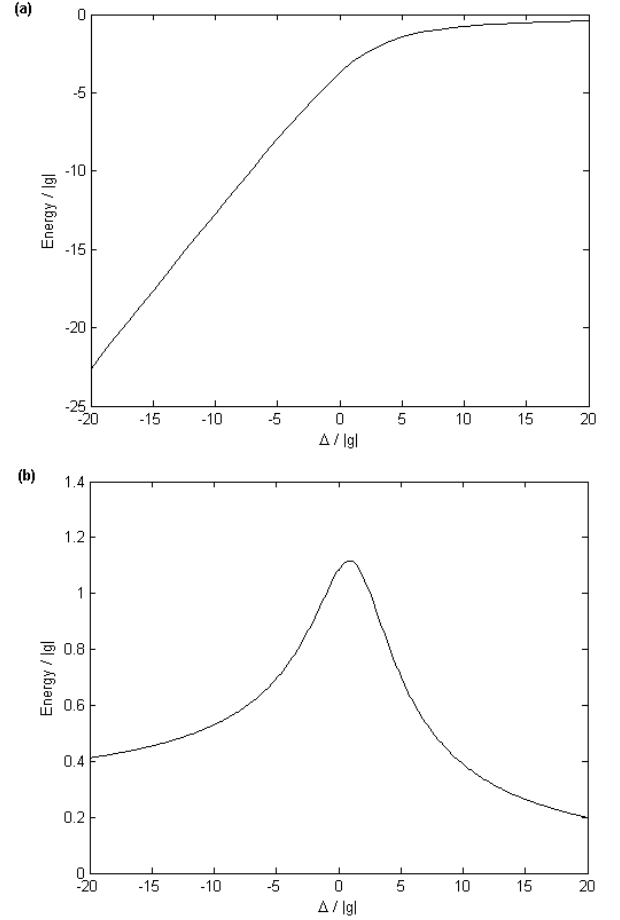} }
\caption{Energy vs. $\Delta $ for a plaquette occupied by three $\uparrow $
and one $\downarrow $ atoms. Other parameters are $t_{da}=1.5|g|$, $%
t_{a}=-0.2|g| $, $t_{d}=0$ (a): Ground state energy. (b): Energy difference
between ground state and first excited state.}
\label{spect31}
\end{figure}

The plaquette occupied by three $\uparrow $ and one $\downarrow $ atoms has
only one type of ground state over the full range of $\Delta $. The
ground state has s-wave symmetry (i.e. it is unchanged under $90^{\circ }$
rotations in the plane of the lattice). The ground state energy is plotted in
FIG. \ref{spect31}(a), and the energy difference between the ground state and
the first excited state is shown in FIG. \ref{spect31}(b). For this figure
the other parameters were $t_{da}=1.5|g|$ and $t_{a}=-0.2|g|$.

The ground state can be represented compactly in the pictorial representation
introduced above. Here we add an additional symbol to represent a single
atom in the $\uparrow $ state. Because the order of the fermionic creation
operators matters, we use $\uparrow $ to represent the left creation
operator and $\Uparrow $ to represent the right creation operator. For
instance, \includegraphics{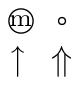} $=d_{1}^{\dagger }a_{3\uparrow
}^{\dagger }a_{4\uparrow }^{\dagger }\left| 0\right\rangle $, whereas %
\includegraphics{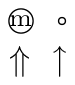} $=d_{1}^{\dagger }a_{4\uparrow }^{\dagger
}a_{3\uparrow }^{\dagger }\left| 0\right\rangle $. \newline

The ground state is: $\left| \psi \right\rangle _{\mathcal{S}}=C_{1}\left|
1\right\rangle _{\mathcal{S}}+C_{2}\left| 2\right\rangle _{\mathcal{S}%
}+C_{3}\left| 3\right\rangle _{\mathcal{S}}$, where

\includegraphics{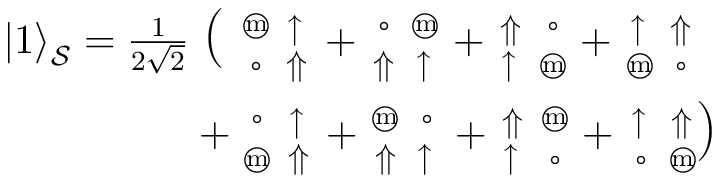}

\includegraphics{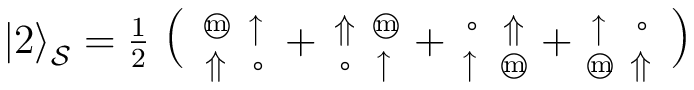}

\includegraphics{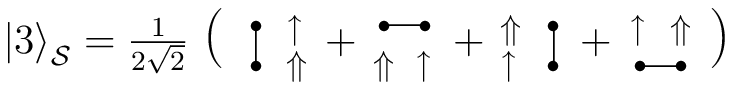}

\begin{figure}[tbp]
{\small \includegraphics[width=229.5pt]{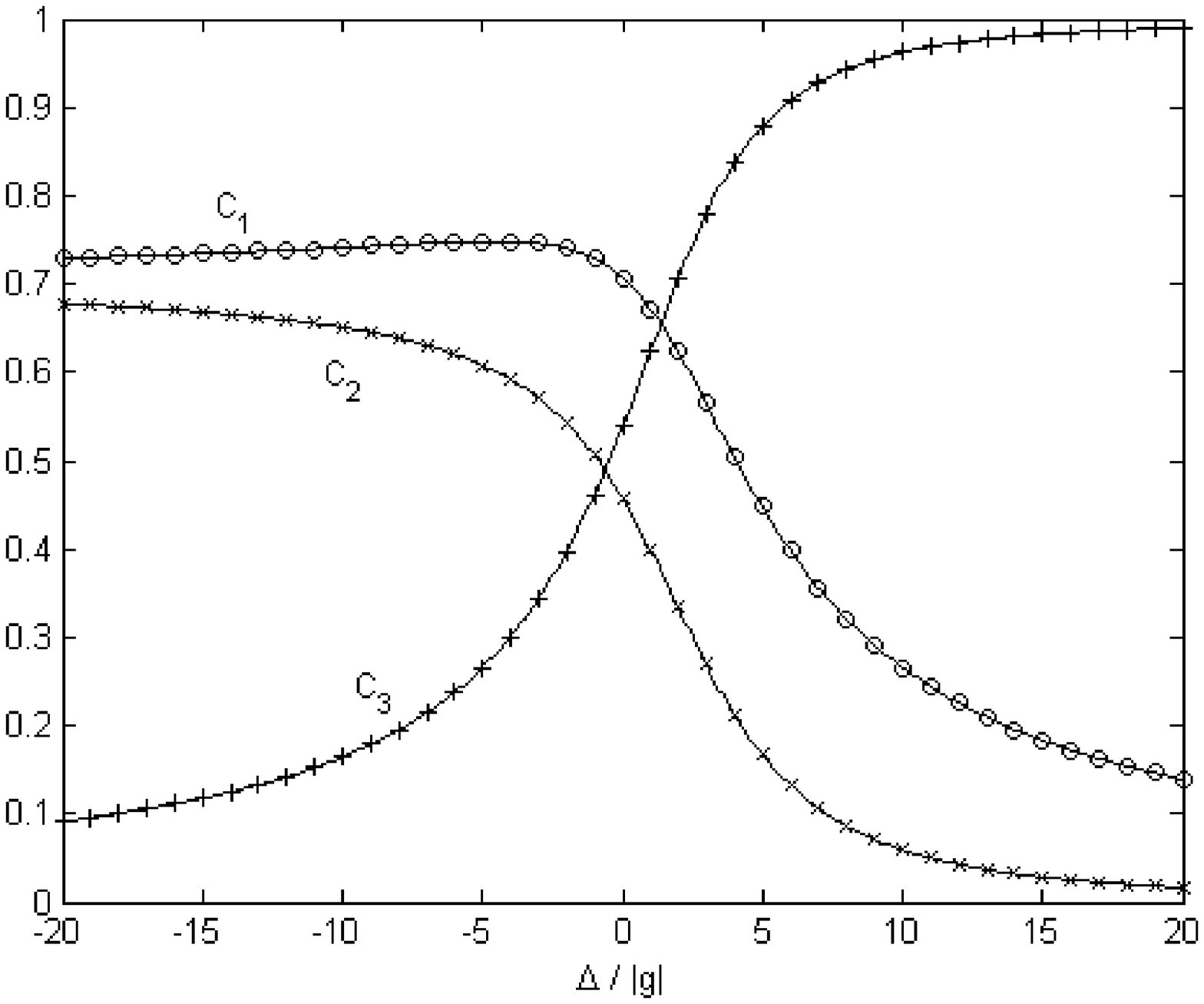} }
\caption{Components of the ground state ($C_{1}$: $\circ $, $C_{2}$: x, $%
C_{3} $: +) vs. $\Delta $ for a plaquette occupied by three $\uparrow $ and
one $\downarrow $ atoms. ($t_{da}=1.5|g|$, $t_{a}=-0.2|g|$, $t_{d}=0$) The
marked datapoints were computed from the full Hamiltonian $H_{eff}$, whereas
the solid lines were computed from the projected Hamiltonian $H_{\mathcal{S}%
} $.}
\label{coeff31}
\end{figure}

The values of the coefficients $C_{1}$, $C_{2}$, $C_{3}$ are shown in FIG.
\ref{coeff31}.

Projected onto the three-dimensional subspace of the full Hilbert space with
basis vectors $\left| 1\right\rangle _{\mathcal{S}}$, $\left| 2\right\rangle
_{\mathcal{S}}$, $\left| 3\right\rangle _{\mathcal{S}}$, $H_{eff}$ (for $%
t_{d} \simeq 0$) is:

{\small
\begin{equation}
H_{\mathcal{S}}=\left(
\begin{array}{ccc}
\Delta & \sqrt{2}\left( t_{a}-t_{da}\right) & -2\sqrt{2}g \\
\sqrt{2}\left( t_{a}-t_{da}\right) & \Delta & 0 \\
-2\sqrt{2}g & 0 & 0
\end{array}
\right)
\end{equation}
}Thus, the ground state of this Hamiltonian is the ground state of $H_{eff}$.
(See solid lines in FIG. \ref{coeff31}.)

It should also be noted that the s-wave ground state for 3 $\uparrow $, 1 $%
\downarrow $ atoms per plaquette is degenerate with the triplet state for 2 $%
\uparrow $ and 2 $\downarrow $ atoms described in the previous section. In
fact, the 2 $\uparrow $, 2 $\downarrow $ triplet state is identical to the
ground state for 3 $\uparrow $ and 1 $\downarrow $ atoms and for 1 $\uparrow $
and 3 $\downarrow $ atoms, except that the triplet $\frac{1}{\sqrt{2}}\left(
a_{i\uparrow }^{\dagger }a_{j\downarrow }^{\dagger }+a_{i\downarrow
}^{\dagger }a_{j\uparrow }^{\dagger }\right) $ is replaced with $%
a_{i\uparrow }^{\dagger }a_{j\uparrow }^{\dagger }$ in the 3 $\uparrow $, 1 $%
\downarrow $ case, and with $a_{i\downarrow }^{\dagger }a_{j\downarrow
}^{\dagger }$ in the 1 $\uparrow $, 3 $\downarrow $ case.

\section{Two atoms per plaquette: one $\uparrow $, one $\downarrow $}

\begin{figure}[tbp]
{\normalsize \includegraphics[width=229.5pt]{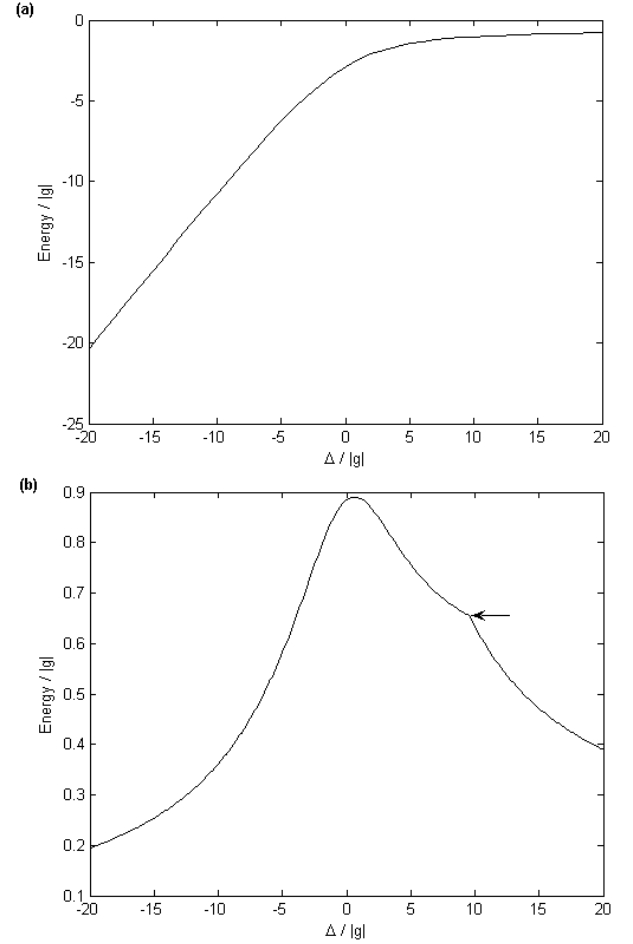} }
\caption{Energy vs. $\Delta $ for a plaquette occupied by one $\uparrow $
and one $\downarrow $ atom. ($t_{da}=1.5|g|$, $t_{a}=-0.2|g|$, $t_{d}=0$.)
(a): Ground state energy (b): Energy difference between ground state and
first excited state. The curve is smooth except at a crossover in the first
excited state (indicated by an arrow).}
\label{spect11}
\end{figure}

When occupied by only a single atom of each spin state, the plaquette has a
single type of ground state for all values of $\Delta $ (for values of the
other parameters within the typical range). This state is symmetric under $%
90^{\circ }$ rotations -- i.e., it has s-wave symmetry. The ground state
energy of this system is plotted in figure \ref{spect11}(a). Figure \ref
{spect11}(b) shows the excitation gap between the ground state and first
excited state. Both these figures assume typical values of $t_{da}$ and $%
t_{a}$ ($t_{da}=1.5|g|$, $t_{a}=-0.2|g|$).

The ground state can be expressed as a vector in a 3-dimensional subspace of
the full Hilbert space. The basis vectors of this subspace (in the pictorial
representation introduced above) are:

\includegraphics{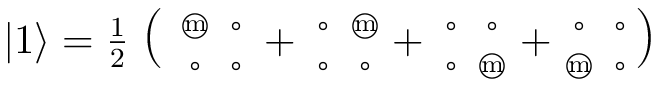}

\includegraphics{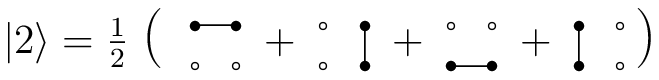}

\includegraphics{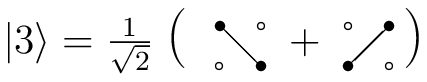}

\begin{figure}[tbp]
{\normalsize \includegraphics[width=229.5pt]{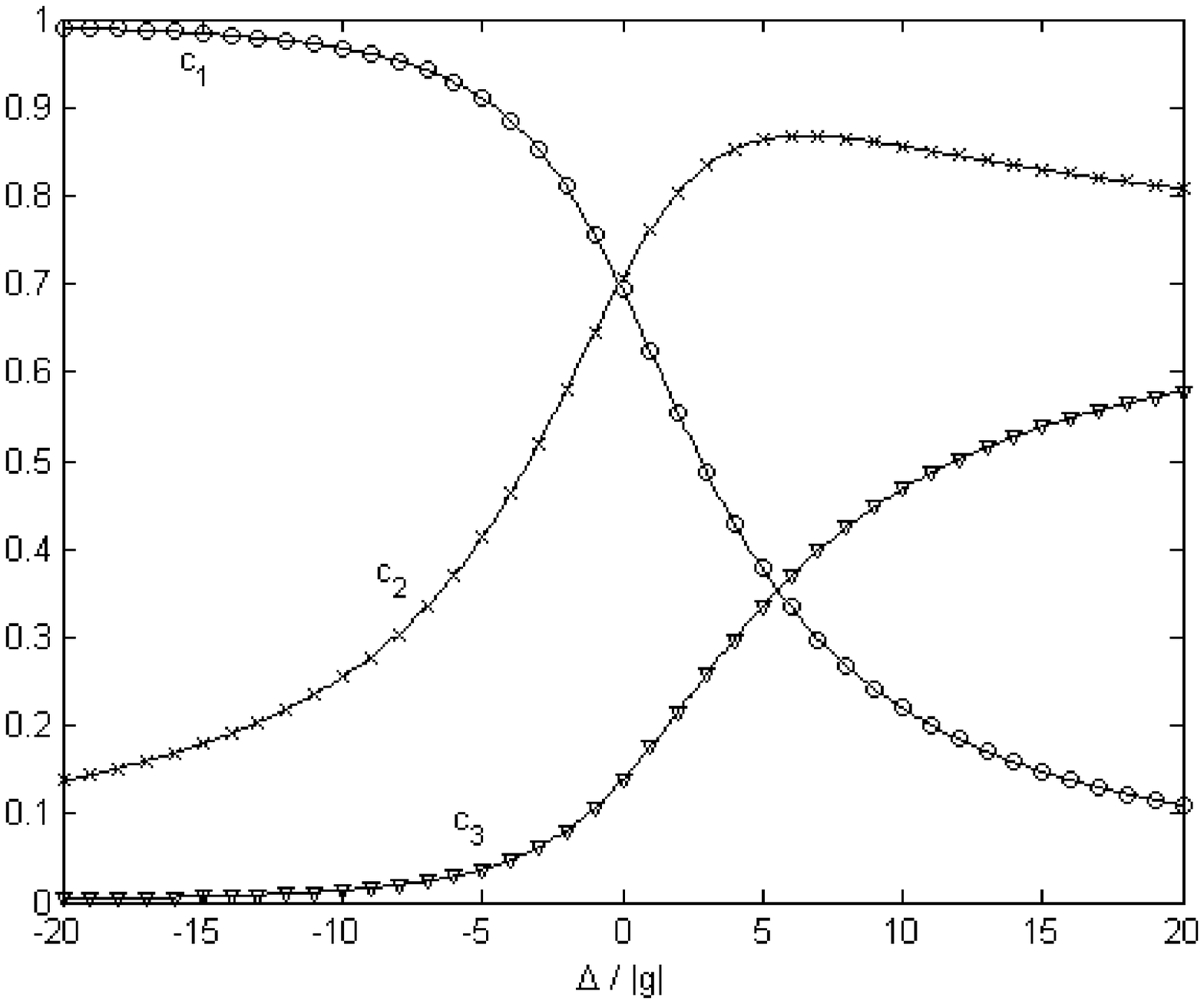} }
\caption{Components of the ground state ($c_{1}$: $\circ $, $c_{2}$: x, $%
c_{3} $: {\protect\footnotesize $\bigtriangledown $}{\protect\small ) vs. $%
\Delta $ for a plaquette occupied by one $\uparrow $ and one $\downarrow $
atom. ($t_{da}=1.5|g|$, $t_{a}=-0.2|g|$, $t_{d}=0$) The marked datapoints
were computed from the full Hamiltonian $H_{eff}$, whereas the solid lines
were computed from the projected Hamiltonian H.}}
\label{coeff11}
\end{figure}

Thus, the ground state is given by: $\left| \psi \right\rangle =c_{1}\left|
1\right\rangle +c_{2}\left| 2\right\rangle +c_{3}\left| 3\right\rangle $,
where the values of the coefficients $c_{1}$, $c_{2}$, $c_{3}$ are shown in
FIG. \ref{coeff11}

Projected onto this subspace, $H_{eff}$ (for $t_{d} \simeq 0$) expressed in
the above basis becomes:
\begin{equation}
H=\left(
\begin{array}{ccc}
\Delta & -2\sqrt{2}g & 0 \\
-2\sqrt{2}g & 0 & 2\sqrt{2}t_{a} \\
0 & 2\sqrt{2}t_{a} & 0
\end{array}
\right)
\end{equation}
Thus, the ground state of this Hamiltonian is the ground state of $H_{eff}$.
(See solid lines on FIG. \ref{coeff11}.)

\section{Three atoms per plaquette:two $\uparrow $, one $\downarrow $}

\begin{figure}[tbp]
{\small \includegraphics[width=7.3cm]{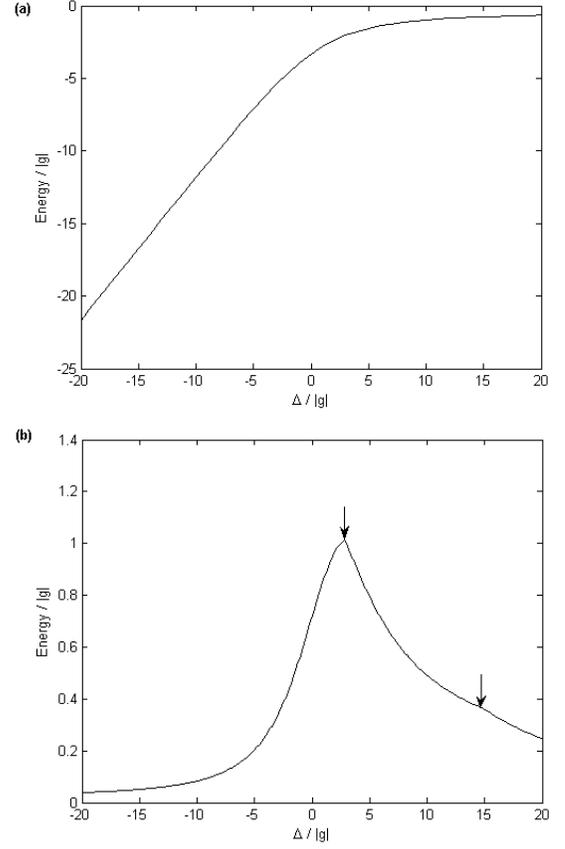} }
\caption{Energy vs. $\Delta $ for a plaquette occupied by two $\uparrow $
and one $\downarrow $ atoms. ($t_{da}=1.5|g|$, $t_{a}=-0.2|g|$, $t_{d}=0$)
(a): Ground state energy (b): Energy difference between ground state and
first excited state. Crossovers in the first excited state (at which points
the curve is not smooth) are indicated by arrows.}
\label{spect21}
\end{figure}

The plaquette with two $\uparrow $ atoms and one $\downarrow $ atom has
three distinct types of ground states for different values of the parameter
$\Delta $ (with the other parameters in the typical range). However, over a
wide range of $\Delta $ around $\Delta =0$ the system is in the same type of
ground state. The ground state of this type is two-fold degenerate. (Hence,
we will refer to this as the ``degenerate state''.) The ground state energy
and the gap between the ground state and first excited state are shown in FIG.
\ref{spect21}. (For the full range of $\Delta $ values shown in the figure,
the system is in the degenerate state.) The degenerate ground states (which
we call $\left| \psi \right\rangle _{+}$ and $\left| \psi \right\rangle _{-}$%
) can be defined in such a way that they are eigenstates of a $90^{\circ }$
rotation in the plane of the plaquette, in which case $\left| \psi
\right\rangle _{\pm }$ gains a factor of $\pm i$ under such a rotation.

The state $\left| \psi \right\rangle _{+}$ can be expressed as a vector in a
particular six-dimensional subspace of the full Hilbert space. We define the
basis vectors of this subspace in the pictorial representation introduced
above. However, because the order of the fermionic creation operators
matters, we use three symbols $\uparrow $, $\Uparrow $, $\wedge $ ($%
\downarrow $, $\Downarrow $, $\vee $) to represent the first, second, and
third creation operator for atoms in the $\uparrow $ ($\downarrow $) state.
E.g., \includegraphics{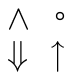} $=a_{4\uparrow }^{\dagger
}a_{3\downarrow }^{\dagger }a_{1\uparrow }^{\dagger }\left| 0\right\rangle $%
. Represented in this way, the six basis vectors are:

\includegraphics{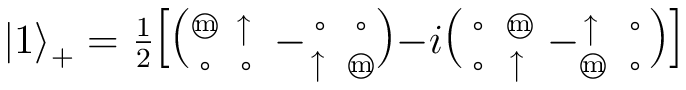}

\includegraphics{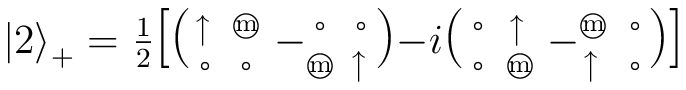}

\includegraphics{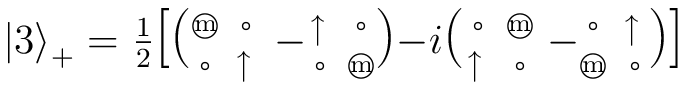}

\includegraphics{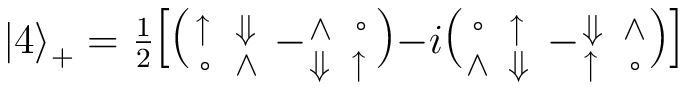}

\includegraphics{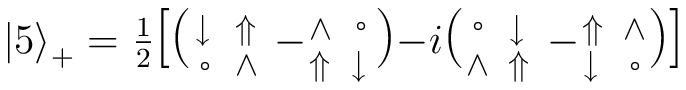}

\begin{figure}[tbp]
{\small \includegraphics[width=229.5pt]{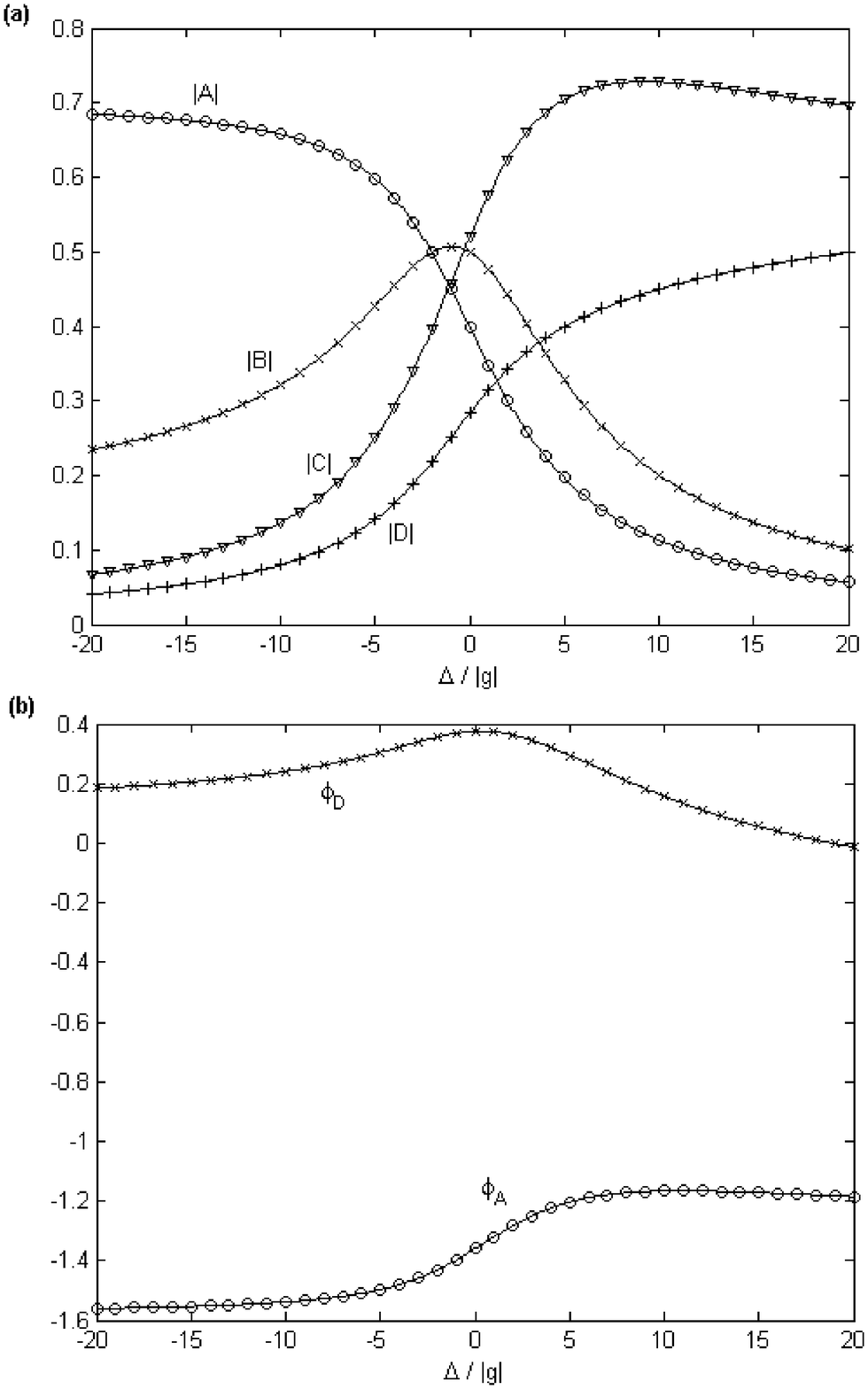} }
\caption{Ground state parameters vs. $\Delta $ for a plaquette occupied by
two $\uparrow $ and one $\downarrow $ atoms. ($t_{da}=1.5|g|$, $t_{a}=-0.2|g|$,
$t_{d}=0$) (a): Amplitudes ($\left| A\right| $: $\circ $, $\left| B\right| $:
x, $\left| C\right| $: {\protect\footnotesize $\bigtriangledown $\protect\small,
$\left|D\right| $: +). (b): Phases ($\protect\phi _{A}$: $\circ $,
$\protect\phi_{D}$: x). The overall phase was chosen to give $B=\left| B\right|
e^{i\protect\pi /4}$ and $C=\left| C\right| e^{i\protect\pi /4}$. The marked
datapoints on (a) and (b) were computed from the full Hamiltonian $H_{eff}$,
whereas the solid lines were computed from the projected Hamiltonian $H_{+}$.%
}}
\label{coeff21}
\end{figure}

\includegraphics{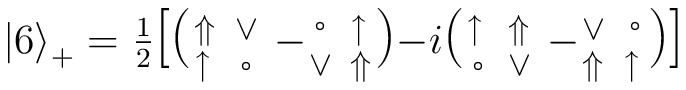}

and the state $\left| \psi \right\rangle _{+}$ is given by:
\begin{eqnarray}
\left| \psi \right\rangle _{+} &=&A\left| 1\right\rangle _{+}+A^{\ast
}\left| 2\right\rangle _{+}+B\left| 3\right\rangle _{+}+C\left|
4\right\rangle _{+}  \nonumber \\
&&+D\left| 5\right\rangle _{+}+D^{\ast }\left| 6\right\rangle _{+}
\end{eqnarray}
for some complex coefficients $A$, $B$, $C$, and $D$. Note that under a $%
90^{\circ }$ clockwise rotation $\left| n\right\rangle _{+}\rightarrow
i\left| n\right\rangle _{+}$ for each $n$, and thus $\left| \psi
\right\rangle _{+}\rightarrow i\left| \psi \right\rangle _{+}$. The state $%
\left| \psi \right\rangle _{-}$ can be expressed as a vector in a
six-dimensional subspace of the full Hilbert space with basis vectors
\begin{equation}
\left| n\right\rangle _{-}=\left| n\right\rangle _{+}^{\ast }
\end{equation}
for $n=1,...,6$. $\left| \psi \right\rangle _{-}$ is given by:
\begin{eqnarray}
\left| \psi \right\rangle _{-} &=&A^{\ast }\left| 1\right\rangle
_{-}+A\left| 2\right\rangle _{-}+B^{\ast }\left| 3\right\rangle _{-}+C^{\ast
}\left| 4\right\rangle _{-}  \nonumber \\
&&+D^{\ast }\left| 5\right\rangle _{-}+D\left| 6\right\rangle _{-}=\left|
\psi \right\rangle _{+}^{\ast }
\end{eqnarray}
The complex coefficients $A$, $B$, $C$, and $D$ can be written as:
\begin{eqnarray}
A &=&\left| A\right| e^{i\phi _{A}},\text{ }D=\left| D\right| e^{i\phi _{D}}
\nonumber \\
B &=&\left| B\right| e^{i\pi /4},\text{ }C=\left| C\right| e^{i\pi /4}
\end{eqnarray}
where $\left| A\right| $, $\left| B\right| $, $\left| C\right| $, $\left|
D\right| $, $\phi _{A}$, and $\phi _{D}$ depend on the parameters of $%
H_{eff} $. The coefficients are determined up to an arbitrary overall phase,
which here was chosen to fix the phases of $B$ and $C$ as shown. ($B$ and $C$
were found to have the same phase.) For typical values of $t_{a}=-0.2|g|$, $%
t_{da}=1.5|g|$, $t_{d}=0$, the values of these parameters (vs. $\Delta $) are
shown in FIG. \ref{coeff21}.

Projected onto the subspace with basis $\left\{ \left| 1\right\rangle
_{+},\left| 2\right\rangle _{+},\left| 3\right\rangle _{+},\left|
4\right\rangle _{+},\left| 5\right\rangle _{+},\left| 6\right\rangle
_{+}\right\} $, $H_{eff}$ (expressed in that basis) is (for $t_{d} \simeq 0$):

\begin{equation}
H_{+}=\left(
\begin{array}{cccccc}
\Delta & t_{da} & t_{a} & -ig & ig & 0 \\
t_{da} & \Delta & -it_{a} & g & 0 & ig \\
t_{a} & it_{a} & \Delta & -2g & g & -ig \\
ig & g & -2g & 0 & -it_{a} & t_{a} \\
-ig & 0 & g & it_{a} & 0 & -t_{a} \\
0 & -ig & ig & t_{a} & -t_{a} & 0
\end{array}
\right)
\end{equation}
The ground state of $H_{+}$ is thus $\left| \psi \right\rangle _{+}$. (See
solid lines in FIG. \ref{coeff21}.) Projected onto the subspace with basis $%
\left\{ \left| 1\right\rangle _{-},\left| 2\right\rangle _{-},\left|
3\right\rangle _{-},\left| 4\right\rangle _{-},\left| 5\right\rangle
_{-},\left| 6\right\rangle _{-}\right\} $, $H_{eff}$ is given by:
\begin{equation}
H_{-}=H_{+}^{\ast }=H_{+}^{T}
\end{equation}
(Note that in this equation $H_{+}$ is still expressed in the basis in which
it was defined above.) Thus the ground state of $H_{-}$ is $\left| \psi
\right\rangle _{+}^{\ast }=\left| \psi \right\rangle _{-}$.

For $\Delta $ far to the negative side ($\Delta <-92.9\left| g\right| $ for
$t_{da}=1.5|g|$, $t_{a}=-0.2|g|$, $t_{d}=0$), the system of two $\uparrow $
atoms and one $\downarrow $ atom on a plaquette has a non-degenerate d-wave
ground state. This state can be expressed as a vector in a 3-dimensional
subspace of the full Hilbert space of this system, with basis vectors:

\includegraphics{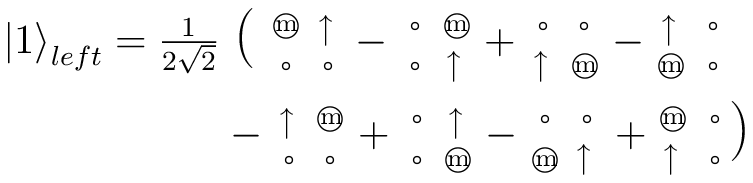}

\includegraphics{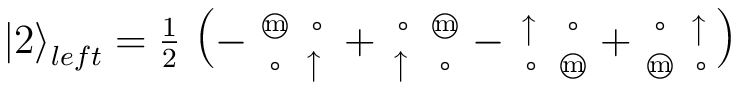}

\includegraphics{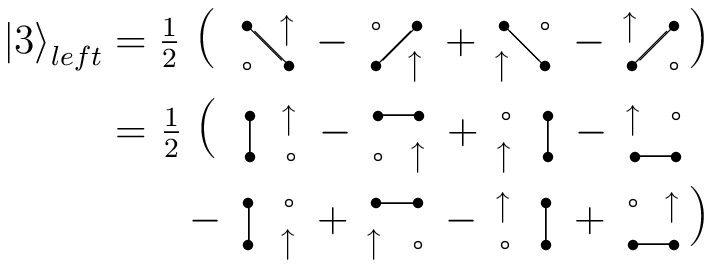}

Note that there is some ambiguity in which spins to group into a singlet and
which to write as $\uparrow $, as evidenced by the two forms given for $%
\left| 3\right\rangle _{left}$ (the first of which has the advantage of all
its terms being orthogonal, but the second of which makes more obvious how
the Hamiltonian connects it to components $\left| 1\right\rangle _{left}$
and $\left| 2\right\rangle _{left}$).

$H_{eff}$ (for $t_{d} \simeq 0$) projected onto this subspace expressed in
the basis $\left\{ \left| 1\right\rangle _{left},\left| 2\right\rangle
_{left},\left| 3\right\rangle _{left}\right\} $ is:
\begin{equation}
H_{left}=\left(
\begin{array}{ccc}
\Delta -t_{da} & -\sqrt{2}t_{a} & -g \\
-\sqrt{2}t_{a} & \Delta & -\sqrt{2}g \\
-g & -\sqrt{2}g & t_{a}
\end{array}
\right)
\end{equation}
Thus, the ground state of this Hamiltonian is the ground state of $H_{eff}$
in the left-most region ($\Delta <-92.9\left| g\right| $).

For $\Delta $ far to the positive side ($\Delta >97.9\left| g\right| $ for $%
t_{da}=1.5|g|$, $t_{a}=-0.2|g|$, $t_{d}=0$), the system has a non-degenerate
s-wave ground state. Furthermore, in this state the ground state wavefunction
and energy are constant for changing $\Delta $. In the pictorial
representation this ground state is given by:

\includegraphics{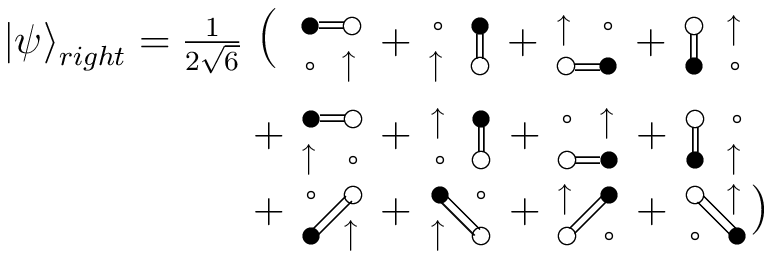}

The energy of this state is $H_{right}=2t_{a}$.

It should be noted that the case of $1 \uparrow$, $2 \downarrow$ atoms on a
plaquette is equivalent to the $2 \uparrow$, $1 \downarrow$ case under exchange
of $\uparrow$ and $\downarrow$ spins.  The Hamiltonian $H_{eff}$ is invariant
under such a spin exchange, except for a change in the sign of $g$.  This is
equivalent to replacing $d^\dagger$ with $-d^\dagger.$  Thus, the
eigenenergies of these two cases are identical, and the eigenstates are
identical except for a change in the sign of the components which include a
dressed molecule.

\section{Summary and discussion}

In the above, we have investigated the ground state properties of the system
with different numbers of spin $\uparrow $ and $\downarrow $ atoms occupying
the four-site plaquette in an optical superlattice. All the other cases can be
reduced to one of the configurations considered above, or to a trivial case,
through the particle-hole exchange.  (Cases where all particles are in the
same spin state are non-interacting, and thus trivial.) For instance, for
five atoms with three spin-$\downarrow $ and two spin-$\uparrow $, one has two
spin-$\uparrow $ and one spin-$\downarrow $ holes in that plaquette. So, the
states are equivalent to those in the case with two spin-$\uparrow $ and one
spin-$\downarrow $ atoms, but with exchange of the parameters $t_{da}$ and
$t_{a}$ in the effective Hamiltonian $H_{eff}$.  The sign of g also changes,
but as noted above this is equivalent to replacing $d^\dagger$ with
$-d^\dagger$.  Thus this change has no effect on the eigenenergies, and the
eigenstates only experience a change in the sign of those components where
the plaquette is occupied by an odd number of dressed molecules.  In
addition, if particle-hole exchange changes the number of atoms by $N$, then
the eigenenergies are shifted by $\frac{N}{2}\Delta$.

From this investigation, we have seen that even on a single plaquette, the
Hamiltonian $H_{eff}$ exhibits a number of different types of ground state
configurations, possessing various forms of rotational symmetry (s-wave,
d-wave, etc.). There are level crossings between these different types of
ground states as the detuning $\Delta $ is varied. The change of the
ground state symmetry from s-wave to d-wave as one scans the parameter $%
\Delta $ from negative to positive regions may be a general feature and not
limited to a single plaquette. For a large lattice, this symmetry change
might correspond to a quantum phase transition from the s-wave to the d-wave
superfluidities \cite{sf}. The states found in this work on a single
plaquette also provide some basic entries for constructing the effective
many-body Hamiltonian for atoms in a quasi-two-dimensional optical lattice
through the contractor renormalization method \cite{core1,core2}. When the
average filling number of the lattice is close to a half with hole doping,
one expects that the basic degrees of freedom from each plaquette are the
ground state configurations specified in Sec. II, the fermionic hole
excitations given by the states in Sec. V, the bosonic hole-pair excitations
specified in Sec. IV, and the bosonic spin excitations given by the states
in Sec. III and II. The effective many-body Hamiltonian will then describe
the interaction between these basic degrees of freedom. So, it is our hope
that the investigation of the single-plaquette physics here will make it
possible to better understand the physics of strongly interacting fermions
on larger lattices.

This work was supported by the NSF awards (0431476), the ARDA under ARO
contracts, and the A. P. Sloan Fellowship.

\end{document}